\documentclass[11pt,a4paper]{article}
\usepackage{jheppub,subfig,graphicx,amsmath}
\usepackage{bbold}
\usepackage{multirow}
\usepackage[bottom]{footmisc}
\usepackage{graphicx,booktabs,array,float}

\title{Encoding  the lattice in the Holography }
\author{Taewon Yuk and Sang-Jin Sin}
\affiliation{Department of Physics, Hanyang University, Seoul 04763, South Korea}
\emailAdd{tae1yuk@gmail.com, sangjin.sin@gmail.com}
\abstract
{One of the most wanted features of holography in its condensed matter physics application is to encode the structure of lattice, which is the most direct data of the material. In this paper, we propose a method to encode the lattice structure by embedding the tight binding data into the Dirac equation in the AdS bulk. We explicitly worked out the idea for the Graphene and Haldane model, and the result shows that some degrees of freedom escape the free-electron on-shell curve, and Green's function loses the pole structure completely. It implies that the electronic structure is not described by the band structure only, which is consistent with what many ARPES data tell us, and it also implies that the system is in non-fermi liquid even for the graphene, which is consistent with recent experiments for the clean graphene.
}
\arxivnumber{arxiv number}

\begin{document}
\maketitle
	  	\section{Introduction}
	  %	\textcolor{red}{%1. description for dual gravity of strong ee interaction. \\
	  	%2. Relating fermion mass $m$ and (which?)interaction strength: $\gamma:=\Delta-\Delta_0=\frac{3}{2}+m-\frac{3}{2}=m$. \\
  	%	3. UV-IR misunderstanding: why there is a lattice?}\\
The physics of strongly interacting many-body systems\cite {Sachdev:2011mz} is one of the most tantalizing subjects of modern physics. However,  one does not have a well-established method to calculate the physical observables of such a system in spite of the long history and strong motivation in its relation to the high Tc superconductivity,  Mott phenomena, spin liquids, Kondo problem, and quantum Hall effect. The holographic duality  \cite{Maldacena:1997re,Witten:1998qj,Gubser:1998bc} has been actively used to discuss the strongly correlated system, where the strong electron-electron interaction is traded by the presence of the semi-classical gravity so that one obtains the solvability of otherwise formidable systems. However, how to include the material structure in this approach is not yet known. For this purpose, the most direct target is encoding the lattice, without which we can not be sure what system we are handling. For example, the holographic method to calculate conductivity is well established  \cite{Donos:2013eha,Donos:2014yya,Hartnoll:2016apf,Faulkner:2009wj,Faulkner:2010da,Blake_2013,Davison_2014,Blake_2015,Ge:2016lyn}. But it has not been clear what system the result is working. Certainly, it is not true that the character of conductivity is the same in all materials. Much of the physical intuition for the physics of condensed matter, even for the strongly correlated systems, comes from electronic structure, and comparing the latter with the data of angle-resolved photoemission spectroscopy (ARPES) is a most basic method,  for which encoding the lattice structure is crucial. Therefore, calculating the fermion spectral function with the encoding lattice is urgently necessary. The holographic fermion spectral function\cite{Iqbal:2009fd,laia2011holographic,Faulkner:2013bna,Oh:2020cym} is well established, but encoding lattice has never been discussed. However, there are a few works, including charge density wave\cite{Ling:2013aya,Balm:2019dxk}, where chemical potential contains the spatial modulation so that it leads to a partial differential equation, which is rather a difficult task to solve even for a one-dimensional lattice. 
	
In this paper, we suggest a method to calculate the electronic structure, including the electron-electron (ee) interaction effect, using holography. The lattice structure is encoded in the tight-binding (TB) data in the absence of the ee interaction, and it contains the Fourier-transformed data of the lattice structure as well as the chemical composition. The idea is to embed the TB data into the Dirac equation in the AdS space and calculate the boundary Green's functions using the established method of holographic fermion.

More specifically, the tight binding result can be written as a multi-component Schr\"odinger equation,  one for each band, that can be written as 
$$\left[i\partial_t-\mathcal{H}(k_x,k_y) \right]\psi=0.$$ 
This is called the 'Dirac equation' due to its multi-component nature, although it is not for a relativistic system. All the lattice information and chemical compositions are included in the  $\mathcal{H}(k_x,k_y)$ in the momentum space. What we are proposing is to embed above equation into the Dirac equation in the Fourier transformed AdS space $$\left[ {\Gamma}^{z}\partial_z-m-i {\Gamma}^{t}\left\{i\mathcal{D}_t-\mathcal{G}\mathcal{H}(k_x,k_y)\right\}\right] \Psi(z, t; k_x,k_y)=0, $$ where $z$ is the coordinate of holographic direction whose 0 is at the boundary of AdS, and ${\Gamma}^{\mu}= e^\mu_a{\Gamma}^{a}$  with flat space gamma matrices ${\Gamma}^{a}$ and vielbein factors   $e^\mu_a$ representing the gravity effect which effectively takes care of the ee interaction. $\mathcal{G}$ is also a factor defined by the vielbein. Notice that we do not consider the anisotropy of the system in this paper but distinguish the time and space since the metric for the finite temperature and density distinguishes the two. The power of our method is that it leads us to an ordinary differential equation, and we can utilize the full power of the tight binding method to encode all the lattice structures as well as the chemical compositions.

We emphasize that the bulk fermion mass $m$ does not really carry a scale but takes care of the interaction strength effectively. This is because $m$ and  the anomalous dimension $\gamma$ of the fermion operator is related by  $$\gamma=\Delta-\Delta_0=\left(\frac{3}{2}-m\right)-1=\frac{1}{2}-m.$$ Here $\Delta$ is the true scaling dimension of the fermion while $\Delta_0$ is the engineering dimension of the latter. From this point of view, the fact that we get fermi liquid for $m=1/2$ \cite{Cubrovic:2010bf} is very natural. Here, we confine ourselves to the two band cases and explicitly work out two of the most studied systems, namely the graphene and the Haldane model. Our result implies that the electronic structure is not described by the band structure only, which is consistent with many  ARPES data,   and it also implies that the system is in non-fermi liquid even for the graphene, which is consistent with recent experiments for the clean graphene \cite{pkim,Seo:2016vks,Lucas:2015sya}.

\section{Tight-binding Hamiltonian}
In this section, we work out ideas for the Graphene and Haldane model. Suppose a tight-binding Hamiltonian as follows:
\begin{align}
	H_{\text{TB}}=\sum_{\boldsymbol{k},\beta,\delta}c_{\boldsymbol{k}\beta}^{\dagger}h_{\text{TB},\beta\delta}(\boldsymbol{k})c_{\boldsymbol{k}\delta},
\end{align}
where $\beta$ and $\gamma$ are the internal physical degree of freedom (e.g. spin, orbital, sublattice, and so forth.). and $c^{\dagger}_{\boldsymbol{k},\beta}(c_{\boldsymbol{k},\beta})$ is fermion creation(annihilation) operator at site $\beta$. For simplicity, we use the matrix notation for $h_{\text{TB}}(\boldsymbol{k})$ deleting the indices. Notice that all the examples we consider in this paper are 2-band models. So, the most general $h_{\text{TB}}$ can be  decomposed into basis  $2 \times 2$ matrices  that are given by the Identity matrix $I_2$ and Pauli matrices $\boldsymbol{\sigma}=\,(\sigma_1,\sigma_2,\sigma_3)$:
\begin{align}
	h_{\text{TB}}(\boldsymbol{k})=h_{0}(\boldsymbol{k})I_2+h_{1}(\boldsymbol{k})\sigma_1+h_{2}(\boldsymbol{k})\sigma_2+h_{3}(\boldsymbol{k})\sigma_3=h_0(\boldsymbol{k})I_2+\boldsymbol{h}\cdot\boldsymbol{\sigma},
\end{align}
where $I_n$ is the $n\times n$ identity matrix. Its eigenvalues are given by
\begin{align}
	E=h_0\pm\sqrt{h_1^2+h_2^2+h_3^2}.
\end{align}
\begin{figure}[H]
	\centering
	\captionsetup{justification=centering}
	\subfloat{\includegraphics[width=6cm]{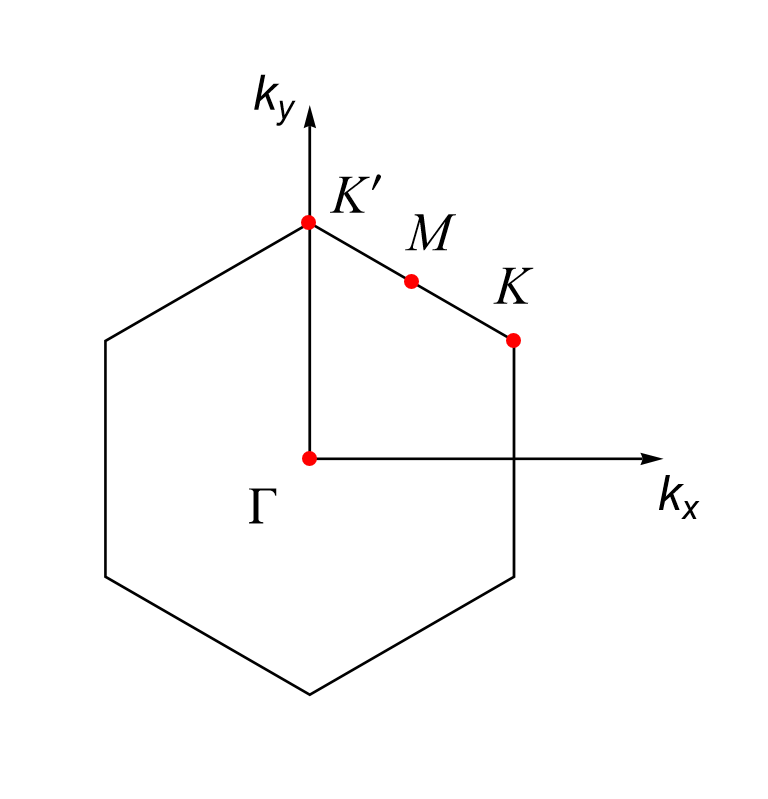}}
	\caption{Brillouin zone of the honeycomb lattice with the high symmetric points. We herein use the $\Gamma$-$K$-$M$-$K'$-$\Gamma$ path.} \label{fig:kpath}
\end{figure}
The honeycomb lattice used in this paper is depicted in Figure \ref{fig:kpath} with the momentum path for the dispersion relation(DR) and spectral density(SD)  in the Brillouin zone(BZ) specified.

\paragraph{Graphene}
Graphene's TB Hamiltonian is given by
\begin{align}
	H_{TB}=-t\sum_{\langle ij \rangle}c_i^{\dagger}c_j,
\end{align}
where $\langle ij \rangle$ means sum of nearest-neighbor(NN) hopping controlled by the parameter $t$. In Figure \ref{fig:Graphene}, we describe the Bravais lattice of the graphene. 
\begin{figure}[H]
	\centering
	\captionsetup{justification=centering}
	\subfloat{\includegraphics[width=6cm]{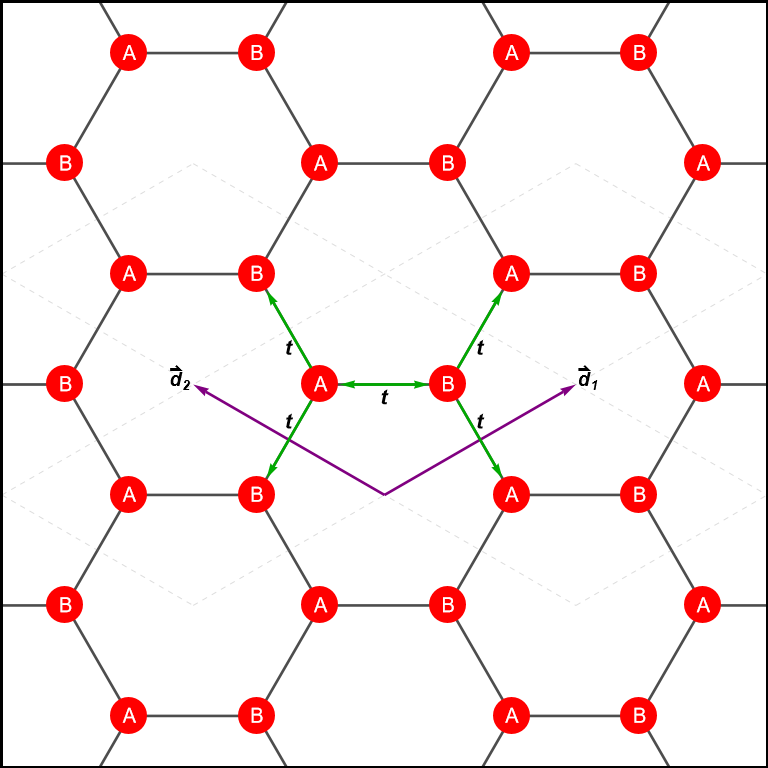}}
	\caption{Bravais lattice of the graphene. The purple line indicates a primitive lattice vector. The gray dashed line means primitive unit cell. The green line represents a fixed unit cell's hopping amplitude $t$.} \label{fig:graphenelattice}
\end{figure}
Here, we define a primitive lattice vector as
\begin{align}
	\vec{d}_1=\frac{\sqrt{3}}{2}a(\sqrt{3},1), \quad \vec{d}_2=\frac{\sqrt{3}}{2}a(-\sqrt{3},1),
\end{align}
where $a$ is a lattice spacing. We consider a fixed unit cell to write down the TB Hamiltonian explicitly. For the fixed unit cell, there are six NN hoppings, which are depicted as the green line in Figure \ref{fig:Graphene}. Then, Hamiltonian can be written as
\begin{align}
	H_{\text{TB}}=-t\sum_{i}\left[ \left(c^{\dagger}_{B,i}+c^{\dagger}_{B,i-\vec{d}_1}+c^{\dagger}_{B,i+\vec{d}_2}\right)c_{A,i}+\left(c^{\dagger}_{A,i}+c^{\dagger}_{A,i+\vec{d}_1}+c^{\dagger}_{A,i-\vec{d}_2}\right)c_{B,i}\right] .
\end{align}
Its Fourier transformation is given by
\begin{align}
	H_{\text{TB}}=-t\sum_{\boldsymbol{k}}\left[ \left(1+e^{i\vec{d}_1\cdot\boldsymbol{k}}+e^{-i\vec{d}_2\cdot\boldsymbol{k}}\right)c^{\dagger}_{B,\boldsymbol{k}}\,c_{A,\boldsymbol{k}}+\left(1+e^{-i\vec{d}_1\cdot\boldsymbol{k}}+e^{i\vec{d}_2\cdot\boldsymbol{k}}\right)c^{\dagger}_{A,\boldsymbol{k}}\,c_{B,\boldsymbol{k}}\right] .
\end{align}
Using the Pauli matrix expression with $\left(c_{A,\boldsymbol{k}},c_{B,\boldsymbol{k}}\right)$ ordering gives the following result:
\begin{align}
	h_{\text{TB}}(\boldsymbol{k})=-t\left[ \left(1+2\cos(\tilde{k}_x)\cos(\tilde{k}_y\right)\sigma_1+2\sin(\tilde{k}_x)\cos(\tilde{k}_y)\sigma_2\right] , \label{eq:graphenetb}
\end{align}
where $\tilde{k}_x=3ak_x/2$, $\tilde{k}_y=\sqrt{3}ak_y/2$.

\paragraph{Haldane model}
  Hamiltonian of the Halande model \cite{PhysRevLett.61.2015} is given by
 \begin{align}
 	H_{TB}=-t\sum_{\langle ij \rangle}c_i^{\dagger}c_j+t_2\sum_{\langle\langle ij \rangle\rangle}e^{-i\nu_{ij}\phi}c_i^{\dagger}c_j+\lambda_{v}\sum_{i}\epsilon_ic_i^{\dagger}c_i.
 \end{align}
 The first term is the same as the Graphene Hamiltonian. The second term indicates a next-nearest-neighbor(NNN) hopping adjusted by internal magnetic flux $\phi$ and hopping parameter $t_2$. Also, the phase factor of the second term depends on the direction of NNN hopping: $$\nu_{ij}=sgn(\hat{d}_i\times\hat{d}_j)_z=\pm1, \quad (i,j)\in\{1,2\}$$ where $\hat{d}_{1,2},$ is the vectors along the directions of NNN hoppings. The last term is on-site energy, which depends on the sublattices:  $\epsilon_i=1$ for A and $\epsilon_i=-1$ for  B. To help the understanding, we also describe the NNN term in Figure \ref{fig:Haldane}.
 \begin{figure}[H]
 	\centering
 	\captionsetup{justification=centering}
 	\subfloat{\includegraphics[width=6cm]{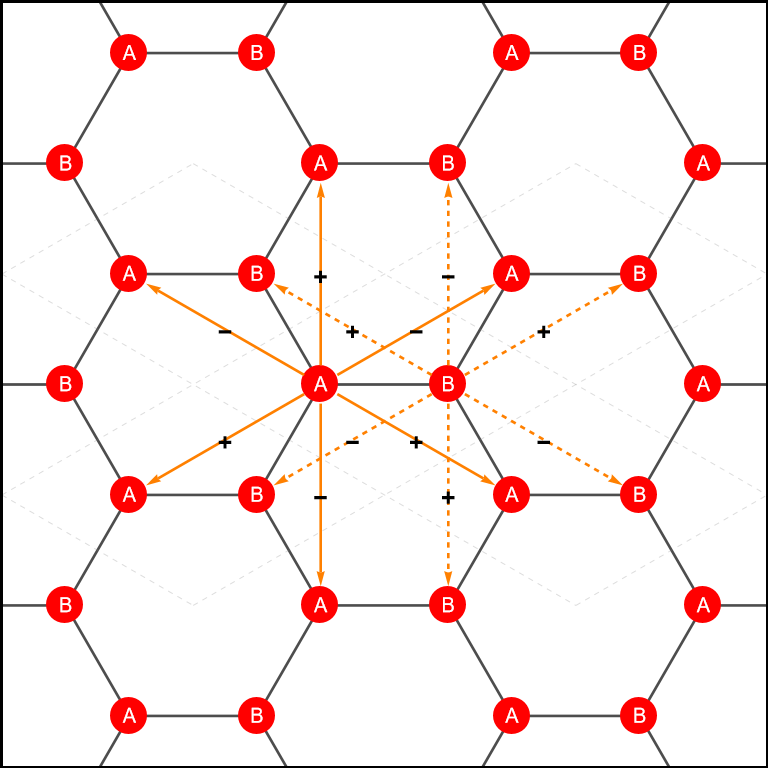}}
 	\caption{Description of the Haldane model in the Bravais lattice. The orange line means NNN hopping in a fixed unit cell. Also, the plus(minus) symbols represents the $\nu_{ij}$ value $+1(-1)$ for each hopping.} \label{fig:haldanelattice}
 \end{figure}
 In figure \ref{fig:haldanelattice}, twelve possible NNN hoppings are drawn in the fixed unit cell with designated $\nu_{ij}$ values. From this, we can write the NNN term as follows:
 \begin{align}
 	t_2&\sum_{i}\Big[ \left(e^{i\phi}c^{\dagger}_{A,i+\vec{d}_1}+e^{-i\phi}c^{\dagger}_{A,i+\vec{d}_1+\vec{d}_2}+e^{i\phi}c^{\dagger}_{A,i+\vec{d}_2}\right)c_{A,i}  \nonumber \\
 	&+   \left(e^{-i\phi}c^{\dagger}_{B,i+\vec{d}_1}+e^{i\phi}c^{\dagger}_{B,i+\vec{d}_1+\vec{d}_2}+e^{-i\phi}c^{\dagger}_{B,i+\vec{d}_2}\right)c_{B,i}+h.c   \Big] .
 \end{align}
 Also, on-site term can be written as 
 \begin{align}
 	\lambda_{v}\sum_{i}\left(c^{\dagger}_{A,i}c_{A,i}-c^{\dagger}_{B,i}c_{B,i}\right).
 \end{align}
 After Fourier transformation, the total Hamiltonian turns out to be
 \begin{align}
 	H_{\text{TB}}= &\sum_{\boldsymbol{k}} \Big[ -t\left(1+e^{i\vec{d}_1\cdot\boldsymbol{k}}+e^{-i\vec{d}_2\cdot\boldsymbol{k}}\right)c^{\dagger}_{B,\boldsymbol{k}}\,c_{A,\boldsymbol{k}}-t\left(1+e^{-i\vec{d}_1\cdot\boldsymbol{k}}+e^{i\vec{d}_2\cdot\boldsymbol{k}}\right)c^{\dagger}_{A,\boldsymbol{k}}\,c_{B,\boldsymbol{k}} \nonumber \\
 	&+2t_2\left(\cos(\phi-\vec{d}_1\cdot\boldsymbol{k})+\cos(\phi+\vec{d}_1\cdot\boldsymbol{k}+\vec{d}_2\cdot\boldsymbol{k})+\cos(\phi-\vec{d}_2\cdot\boldsymbol{k})\right)c^{\dagger}_{A,\boldsymbol{k}}\,c_{A,\boldsymbol{k}} \nonumber \\
 	&+2t_2\left(\cos(\phi+\vec{d}_1\cdot\boldsymbol{k})+\cos(\phi-\vec{d}_1\cdot\boldsymbol{k}-\vec{d}_2\cdot\boldsymbol{k})+\cos(\phi+\vec{d}_2\cdot\boldsymbol{k})\right)c^{\dagger}_{B,\boldsymbol{k}}\,c_{B,\boldsymbol{k}} \nonumber \\
 	&+\lambda_{v}\left(c^{\dagger}_{A,\boldsymbol{k}}c_{A,\boldsymbol{k}}-c^{\dagger}_{B,\boldsymbol{k}}c_{B,\boldsymbol{k}}\right)\Big] .
 \end{align}
 Using the same decomposition to graphene case, $h_{\text{TB}}(\boldsymbol{k})$ can be expressed as
 \begin{align}
 	h_{\text{TB}}(\boldsymbol{k})=&\,2t_2\cos(\phi)\left(2\cos(\tilde{k}_x)\cos(\tilde{k}_y)+\cos(2\tilde{k}_y)\right)I_2-t\left(1+2\cos(\tilde{k}_x)\cos(\tilde{k}_y)\right)\sigma_1 \nonumber \\
 	&-2t\sin(\tilde{k}_x)\cos(\tilde{k}_y)\sigma_2+\left(\lambda_{v}+2t_2\sin(\phi)\left(2\cos(\tilde{k}_x)\sin(\tilde{k}_y)-\sin(2\tilde{k}_y)\right)\right)\sigma_3. \label{eq:haldanetb}
 \end{align}
	
\section{Embedding the  Graphene lattice into holography}
Now, we turn to embed the tight-binding Hamiltonian into holography. 
We start with the action in AdS$_4$. 
\begin{align}
	S_{tot}=&S_{\psi}+S_{bdy}+S_{g,A}, \label{eq:actioni} \\
	S_{\psi}=&\int d^4x\sqrt{-g}\left[i\bar{\psi}(\Gamma^{\mu}D_{\mu}-m)\psi\right], \label{eq:bulkaction} \\
	S_{bdy}=& i\int d^3x\sqrt{-h}\bar{\psi}\psi , \label{eq:actionf} \\
	S_{g,A}=& \int d^4x\sqrt{-g} \left(R+\frac{6}{L^2}-\frac{1}{4}F^2_{\mu\nu}\right).
\end{align}
The background fields are given by
\begin{gather}
	ds^2=-\frac{f(r)}{r^2}dt^2+\frac{dx^2+dy^2+dz^2}{r^2}+\frac{dr^2}{r^2f(r)}, \nonumber \\ f(r)=1-\left(\frac{r}{r_H}\right)^3-\frac{\mu^2r^3}{4r_H}\left(1-\frac{r}{r_H}\right), \\
	A_{\nu}dx^\nu=\mu\left(1-\frac{r}{r_H}\right)dt, \quad r_H=\frac{6}{4\pi T+\sqrt{16 \pi^2 T^2 + 3 \mu^2}},
	\label{eq:background}
\end{gather}
where $r$ is the coordinate of the inverse radius and underlined indices represent tangent space ones.

The Dirac equation on the curved spacetime is written as
\begin{align}
	(\Gamma^{\mu} D_{\mu}-m)\psi=0,
\end{align}
where $D_\mu=\partial_\mu+\frac{1}{4}\omega_{\underline{\nu}\underline{\lambda},\mu}\Gamma^{\underline{\nu}\underline{\lambda}}-iq A_\mu$, $\omega_{\underline{\nu}\underline{\lambda},\mu}$ is the spin connection. 
The 	gamma matrices are given by 
\begin{align}
	\Gamma^{\underline{t}}=-i\sigma_1 \otimes \sigma_0, \quad \Gamma^{\underline{x}}=\sigma_2 \otimes \sigma_1, \quad \Gamma^{\underline{y}}=\sigma_2 \otimes \sigma_2, \quad \Gamma^{\underline{z}}=\sigma_2 \otimes \sigma_3, \quad \Gamma^{\underline{r}}=\sigma_3 \otimes \sigma_0, \label{eq:rep} 	\end{align}
Substituting 
$$\psi=(-gg^{rr})^{-1/4} e^{-i\omega t+i k_xx+i k_yy+i k_zz}\zeta(r),$$
into the Dirac equation, we   get a simplified    one\cite{Liu:2009dm} without the spin connection term:
\begin{align}
	\left[\Gamma^{r}\partial_r-m-{i}\left((\omega+q A_t) e^{t}_{\underline{t}}\Gamma^{\underline{t}}-(k_x e^{x}_{\underline{x}}\Gamma^{\underline{x}} +k_y e^{y}_{\underline{y}}\Gamma^{\underline{y}} +k_z e^{z}_{\underline{z}}\Gamma^{\underline{z}})\right)\right] \zeta=0, \label{eq:diraceq}
	%\\
	%		\left[\Gamma^{r}\partial_r-m-i\left\{\Gamma^{\underline{t}}\left((\omega+q A_t) e^{t}_{\underline{t}}+\Gamma^{\underline{t}}(k_x e^{x}_{\underline{x}}\Gamma^{\underline{x}} +k_y e^{y}_{\underline{y}}\Gamma^{\underline{y}}+k_z e^{z}_{\underline{z}}\Gamma^{\underline{z}})\right)\right\} \right] \zeta=0,
\end{align}
where  $e^{\mu}_{\underline{\mu}}$ means vielbein. Since applying anisotropic geometry (e.g., p- and d- wave superconductor.) still needs to be clarified, we only consider an \textit{isotropic} spatial geometry. Then, we can rewrite the above equation as follows:
	\begin{gather}
	%		\left[\Gamma^{r}\partial_r-m-ie^{t}_{\underline{t}}\left\{\Gamma^{\underline{t}}\left(\omega+q A_t+\frac{e^{x}_{\underline{x}}}{e^{t}_{\underline{t}}}\Gamma^{\underline{t}}(k_x \Gamma^{\underline{x}} +k_y \Gamma^{\underline{y}} +k_z \Gamma^{\underline{z}})\right)\right\}\right] \zeta=0, \\
	%		\left[\Gamma^{r}\partial_r-m-ie^{t}_{\underline{t}}\left\{\Gamma^{\underline{t}}\left(\omega+q A_t+\sqrt{-\frac{g_{tt}}{g_{ii}}}(k_x \Gamma^{\underline{t}}\Gamma^{\underline{x}} +k_y \Gamma^{\underline{t}}\Gamma^{\underline{y}}+k_z \Gamma^{\underline{t}}\Gamma^{\underline{z}})\right)\right\}\right] \zeta=0, \\
	%		\left[\Gamma^{r}\partial_r-m-ie^{t}_{\underline{t}}\left\{\Gamma^{\underline{t}}\left(\omega+q A_t-\mathcal{G}(r) \, \boldsymbol{k} \cdot \boldsymbol{\alpha}\right)\right\}\right] \zeta=0, \\
	\left[\Gamma^{r}\partial_r-m-i\left\{\Gamma^{t}\left(\omega+q A_t-\mathcal{G}(r) \, \boldsymbol{k} \cdot \boldsymbol{\alpha}\right)\right\}\right] \zeta=0, \label{eq:simeq}
\end{gather}
where
\begin{gather}
	\mathcal{G}(r)=\sqrt{-\frac{g_{tt}}{g_{ii}}}=\sqrt{f(r)}, \quad \boldsymbol{k}=(k_x,k_y,k_z), \\
	\boldsymbol{\alpha}=-\Gamma^{\underline{t}}(\Gamma^{\underline{x}}, \, \Gamma^{\underline{y}}, \, \Gamma^{\underline{z}})=-\sigma_3 \otimes(\sigma_1,\sigma_2,\sigma_3).
\end{gather}
Notice that  inside of $\Gamma^{\underline{t}}$ in \eqref{eq:simeq}, one can interpret $\boldsymbol{k} \cdot \boldsymbol{\alpha}$ as Dirac Hamiltonian $\mathcal{H}(\boldsymbol{k})$.
\begin{gather}
	%		\left[\Gamma^{r}\partial_r-m-i\Gamma^{t}\left\{(\omega+q A_t)-\mathcal{G}\mathcal{H}(\boldsymbol{k})\right\}\right] \zeta=0, \\
	\left[\Gamma^{r}\partial_r-m-i\Gamma^{t}\left\{i\mathcal{D}_t-\mathcal{G}\mathcal{H}(\boldsymbol{k})\right\}\right] \zeta=0,
	\quad \hbox{ with } i\mathcal{D}_t=\left(\omega+q A_t\right).  \label{Diraceq}
\end{gather}
So far, we just rewrote the Dirac equation in the curved spacetime. 

Now, to guide the method of embedding, we rewrite Dirac Hamiltonian in the following form:
\begin{align}
	\mathcal{H}(\boldsymbol{k})=-\sigma_3 \otimes (k_x\sigma_1+k_y\sigma_2+k_z\sigma_3).
\end{align}
It is well known that in the holographic calculation of the fermion Green's function, half of the bulk degrees of freedom is projected out, and half is determined by the boundary action. Depending on the choice, we have standard quantization or alternative quantization. In our case,  we will see that only the right factor of the tensor product,  the $2 \times 2$ matrix, reflects Green's function structure of the boundary theory. So, we must generalize the 2 by 2 tight-binding Hamiltonian to 4 by 4 Hamiltonian to embed the tight binding result in the real system to the holographic bulk theory. Naturally, we would generalize by $h_{\text{TB}}(\boldsymbol{k}) \to  -\sigma_3\otimes h_{\text{TB}}(\boldsymbol{k})$. However, the general tight binding method would contain the potential term as well as the kinetic part. For example, in the Haldane model,  $h_{\text{TB}}(\boldsymbol{k})$ has a term proportional to $I_2$, which is absent in the usual Dirac Hamiltonian. Therefore we treat such term as a part of the potential term, while we treat the $\boldsymbol{k}\cdot\boldsymbol{\sigma}$ part as the kinetic part of the Hamiltonian. Therefore, we propose that we embed the tight-binding Hamiltonian to the AdS Dirac equation by 
\begin{align}
	\omega +qA_t &\to \omega+qA_t  - h_0, \nonumber \\
	h_{\text{TB}}(\boldsymbol{k}) &\to  -\sigma_3\otimes h_{\text{TB}}(\boldsymbol{k}).	  \label{eq:embedding1}
\end{align}
We call this the standard embedding. In Appendix B, we describe another possibility where we do not distinguish the kinetic and potential parts as possible choices.  
%Namely, to $h_{\text{TB}}(\boldsymbol{k})$ where we call it as $\textbf{Embedding 1}$:
%	\begin{align}
	%		\textbf{Embedding 1}\text{:} \quad \mathcal{H}_{\text{TB}}(\boldsymbol{k})&:=-\mathcal{G}^{-1}h_0\,I_4-\sigma_3 \otimes \left(\boldsymbol{h}\cdot\boldsymbol{\sigma}\right). \label{eq:embedding1}
	%	\end{align}
%	Also, we recognize there is another embedding given by
%	\begin{align}
	%		\textbf{Embedding 2}\text{:} \quad \mathcal{H}_{\text{TB}}(\boldsymbol{k})&:=-h_0\,I_4-\sigma_3 \otimes \left(\boldsymbol{h}\cdot\boldsymbol{\sigma}\right).
	%	\end{align}
%	Here, the essential difference is the presence of $\mathcal{G}$. In this paper, we choose the first one, which describes a reason.
%
If one wants to keep the prescription to have the same form of \eqref{Diraceq}, one may express our embedding proposal as follows. 
%	Since \eqref{eq:embedding1} shows a minus sign in the front of $\boldsymbol{\sigma}$, we also consider a minus sign of $h_0$. Precisely, we embed $2\times2$ TB Hamiltonian to $4\times4$ Bulk Hamiltonian as follows:
\begin{align}
	h_{\text{TB}}(\boldsymbol{k}) =	h_0I_2+\boldsymbol{h} \cdot \boldsymbol{\sigma} \,\, \rightarrow \,\, 	{\cal H}_{\text{TB}}(\boldsymbol{k}) =\mathcal{G}^{-1}h_0I_4+\boldsymbol{h} \cdot \boldsymbol{\alpha}, \\
	\left[\Gamma^{r}\partial_r-m-i\Gamma^{t}\left\{i\mathcal{D}_t-\mathcal{G}\mathcal{H}_{\text{TB}}(\boldsymbol{k})\right\}\right] \zeta=0. 
	% \\
	%			\mathcal{H}_{\text{TB}}(\boldsymbol{k}):=-\mathcal{G}^{-1}h_0\,I_4-\sigma_3 \otimes \left(\boldsymbol{h}\cdot\boldsymbol{\sigma}\right).
\end{align}
It is important  that the coupling of the  fermion and gravity not only comes from $\mathcal{G}(r)$ but also $\Gamma^{t}$ and $\Gamma^{r}$, because the latter contains gravity information in 
\begin{align}
	e_{\underline{t}}^{t}\,\Gamma^{\underline{t}}, \quad e_{\underline{r}}^{r}\,\Gamma^{\underline{r}}.
\end{align}

Now, to see the effect of the proposal,  we show the resulting  Green's function, the derivation of which is described in Appendix A. For the  pure $AdS$ background, 
\begin{align}
	\left[\Gamma^{r}\partial_r-m-i\Gamma^{t}\left\{\omega-h_0-\boldsymbol{h}\cdot\boldsymbol{\alpha}\right\}\right] \zeta=0,
\end{align}
where $\mathcal{G}(r)=1$. From this Dirac equation, we construct a flow equation, which describes the bulk dynamics of fermionic Green's function.   
\begin{align}\textsc{}
	\partial_r\xi(r)+(\omega-h_0-\boldsymbol{h} \cdot \boldsymbol{\sigma})+\frac{2m}{r}\xi(r)+\xi(r)(\omega-h_0+\boldsymbol{h} \cdot \boldsymbol{\sigma})\xi(r)=0, \label{eq:floweq}
\end{align}
where
\begin{align}
	\xi(r):=
	\begin{pmatrix}
		\xi_{11}(r) & \xi_{12}(r) \\
		\xi_{21}(r) & \xi_{22}(r)
	\end{pmatrix}.
\end{align}
This equation was introduced in \cite{Yuk:2022lof} by generalizing the one given in  \cite{Liu:2009dm}. 
With the infalling boundary condition, $$\xi(r_H)=iI_2,$$ we can get  the  analytic solution of $\mathbb{G}(r)$, 
\begin{align}
	\xi(r)=\frac{1}{\sqrt{\boldsymbol{h}^2-(\omega-h_0)^2}}\frac{K_{m+\frac{1}{2}}(\sqrt{\boldsymbol{h}^2-(\omega-h_0)^2}\,r)}{K_{m-\frac{1}{2}}(\sqrt{\boldsymbol{h}^2-(\omega-h_0)^2}\,r)}
	\begin{pmatrix}
		\omega-h_0-h_3 & -h_1+ih_2 \\
		-h_1-ih_2 & \omega-h_0+h_3
	\end{pmatrix}.
\end{align}
Here, $K_\nu$ is the modified Bessel function of the second type. For the boundary Green's function is defined by 
\begin{align}
	G_R:=\lim_{r\rightarrow 0}r^{2m}\xi(r),
\end{align}
the full matrix solution can be expressed as
\begin{align}
	G_R&=4^m\frac{\Gamma(\frac{1}{2}+m)}{\Gamma(\frac{1}{2}-m)}\frac{\omega-h_0(\boldsymbol{k})-\boldsymbol{h}(\boldsymbol{k})\cdot\boldsymbol{\sigma}}{(\boldsymbol{h}(\boldsymbol{k})^2-(\omega-h_0(\boldsymbol{k}))^2)^{\frac{1}{2}+m}} \nonumber \\
	&=4^m\frac{\Gamma(\frac{1}{2}+m)}{\Gamma(\frac{1}{2}-m)}\frac{\omega-h_{\text{TB}}(\boldsymbol{k})}{(\boldsymbol{h}(\boldsymbol{k})^2-(\omega-h_0(\boldsymbol{k}))^2)^{\frac{1}{2}+m}}. \label{eq:analyticGF}
\end{align}
Notice that although the interaction heavily modifies Green's function, its matrix structure is the same as TB Hamiltonian's, as we mentioned above. Consequently, the topology of the configuration does not change because the overall scalar multiplication gives just a gauge transformation of the Berry potential.
In other words,  the holographic treatment of the fermion gives a Green's function, which gives interaction-induced dressing to the bare non-interacting fermion Green's function. However, such dressing does not change the topology of the many body configurations as advocated before in \cite{PhysRevB.86.165116}. Since the 2-band model can be expressed in terms of the coefficients of Pauli matrices, by plugging the coefficients into \eqref{eq:analyticGF}, we get Green's function containing both lattice structure information as well as the ee interaction.
Also, our method does not involve solving PDE since we work in the momentum space where the Fourier-transformed data of the lattice structure is already encoded in the tight-binding Hamiltonian. This is the difference and merit of our method compared with other works \cite{Ling:2013aya,Balm:2019dxk}.
		
	\section{A few applications}
	In this section, we  apply our method to the graphene and Haldane model and show the dispersion relation(DR) and spectral density(SD) defined by 
	\begin{align}
		E=h_0\pm\sqrt{h_1^2+h_2^2+h_3^2}, \quad A=\text{Im}\,\text{Tr}(G_R).
	\end{align}
	\paragraph{Graphene}
	First, we remind our embedded Green's function for the graphene case, which is given by
	\begin{align}
		G_R=&4^m\frac{\Gamma(\frac{1}{2}+m)}{\Gamma(\frac{1}{2}-m)}\frac{\omega-h_{\text{TB}}(\boldsymbol{k})}{(\boldsymbol{h}(\boldsymbol{k})^2-(\omega-h_0(\boldsymbol{k}))^2)^{\frac{1}{2}+m}}, \\
		h_{\text{TB}}(\boldsymbol{k})=&-t\left[ \left(1+2\cos(\tilde{k}_x)\cos(\tilde{k}_y\right)\sigma_1+2\sin(\tilde{k}_x)\cos(\tilde{k}_y)\sigma_2\right].
	\end{align}
	\begin{figure}[H]
		\centering
		\captionsetup{justification=centering}
		\subfloat[Dispersion Relation of the Graphene]{\includegraphics[width=4.5cm]{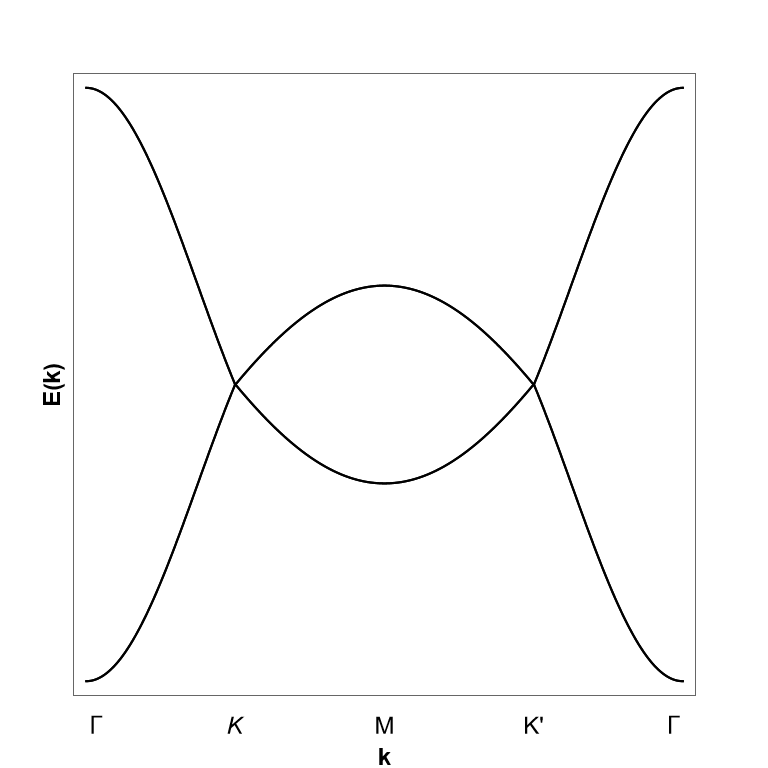}}
		\subfloat[Spectral Density of the Graphene]{\includegraphics[width=4.5cm]{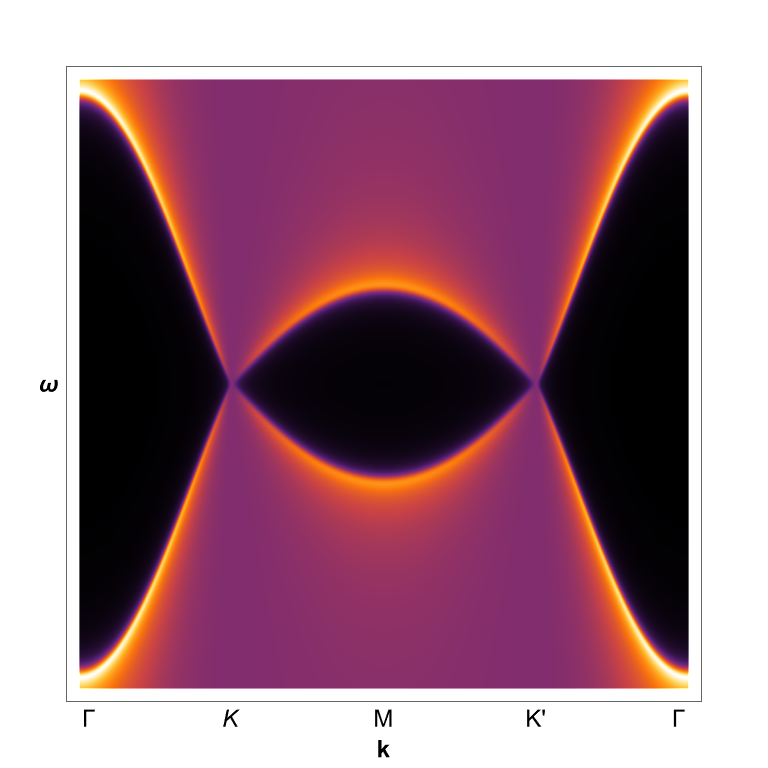}}
		\caption{Comparison between the dispersion relation and the spectral density. The dispersion curves of (a) and (b) are well-matched.} \label{fig:Graphene}
	\end{figure}
	As we can see in Figure \ref{fig:Graphene}, the edge location of the spectral density is the same as the ordinary tight binding result. However, our calculation shows that some degrees of freedom escape to the region allowed by the causality so that the spectral function becomes fuzzy. Such character is very different from the usual perturbative calculation of Green's function, where the interaction effect is limited to the shift of the pole position and broadening of the spectral curve. In this sense, the result is highly non-local in momentum space. This can be understood from the analytic behavior of Green's function given in \eqref{eq:analyticGF}. Our Green's function has branch-cut type singularity while the usual weakly interacting system's Green's function has pole type with complex-valued self-energy. Since the Green's function lost the pole structure completely,  our result show that the system is non-fermi liquid. Notice that it is known that the clean graphene becomes strongly interacting, showing the anomalous transport \cite{pkim}, and our result supports it from a different point of view.
	 
\begin{figure}[H]
	\centering
	\captionsetup{justification=centering}
	\subfloat[$(\phi,\frac{\lambda_{v}}{3\sqrt{3}t_2})=(-\frac{\pi}{2},1.75)$]{\includegraphics[width=8cm]{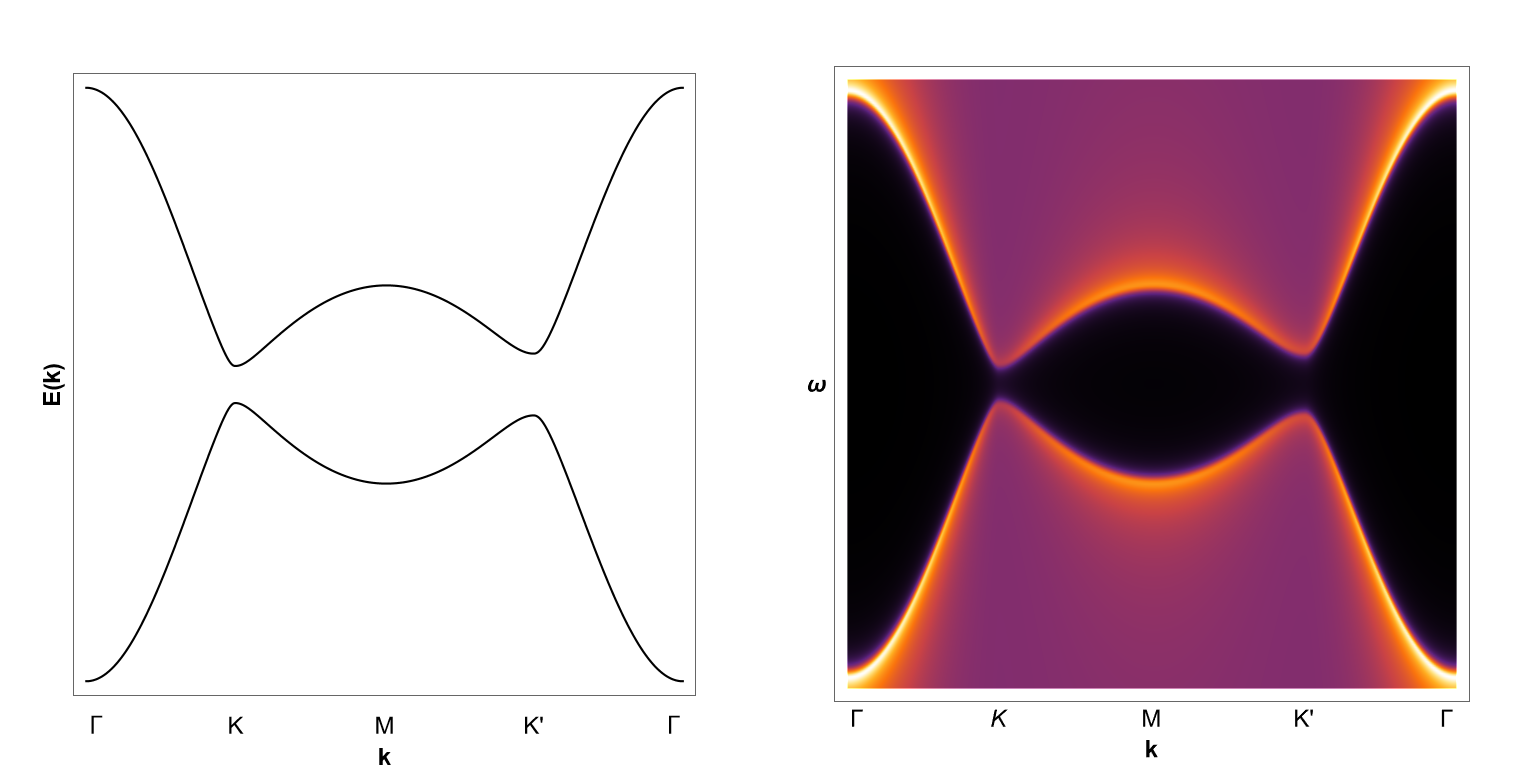}}
	\subfloat[$(\phi,\frac{\lambda_{v}}{3\sqrt{3}t_2})=(-\frac{\pi}{2},1)$]{\includegraphics[width=8cm]{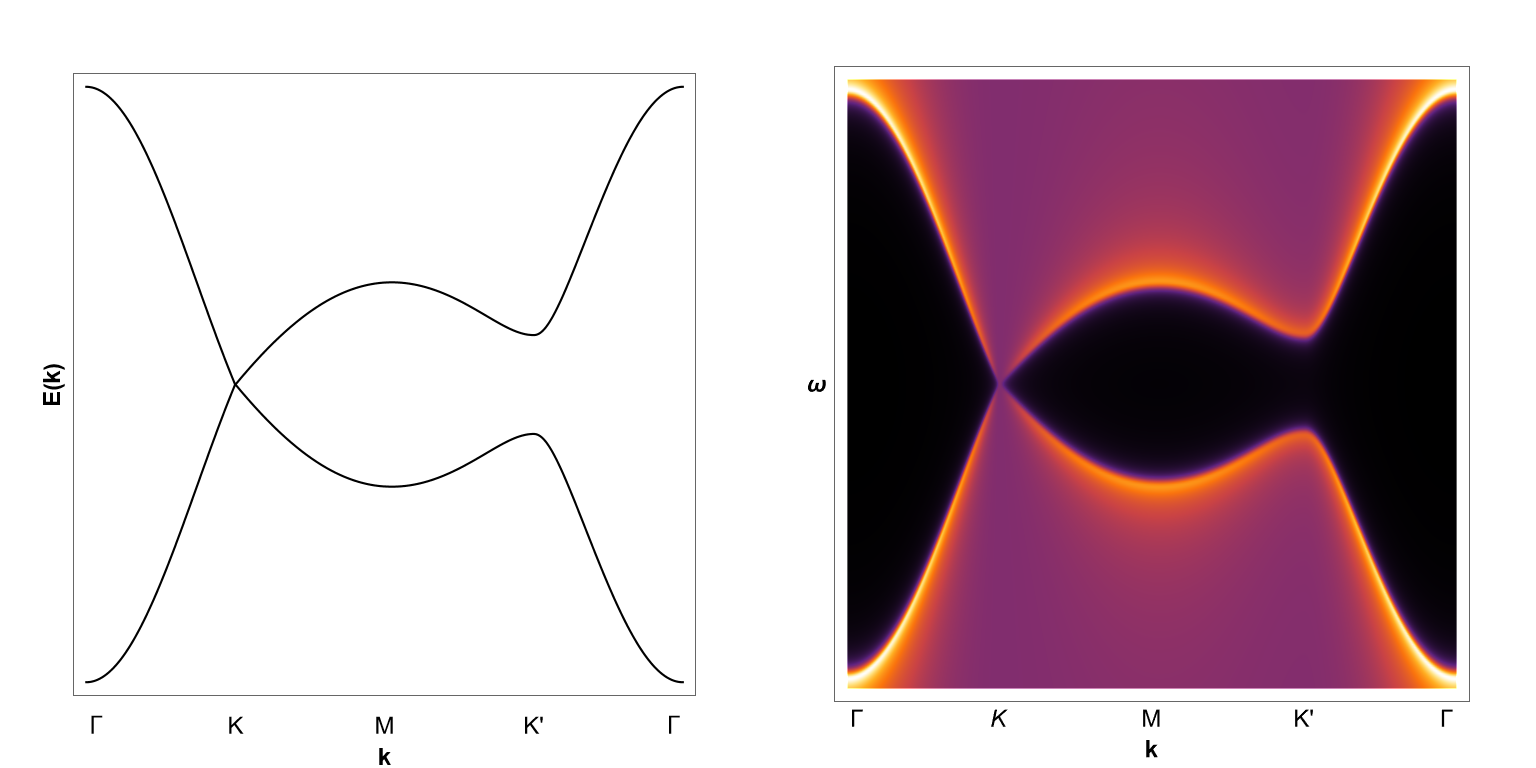}} \\
	\subfloat[$(\phi,\frac{\lambda_{v}}{3\sqrt{3}t_2})=(-\frac{\pi}{2},0.25)$]{\includegraphics[width=8cm]{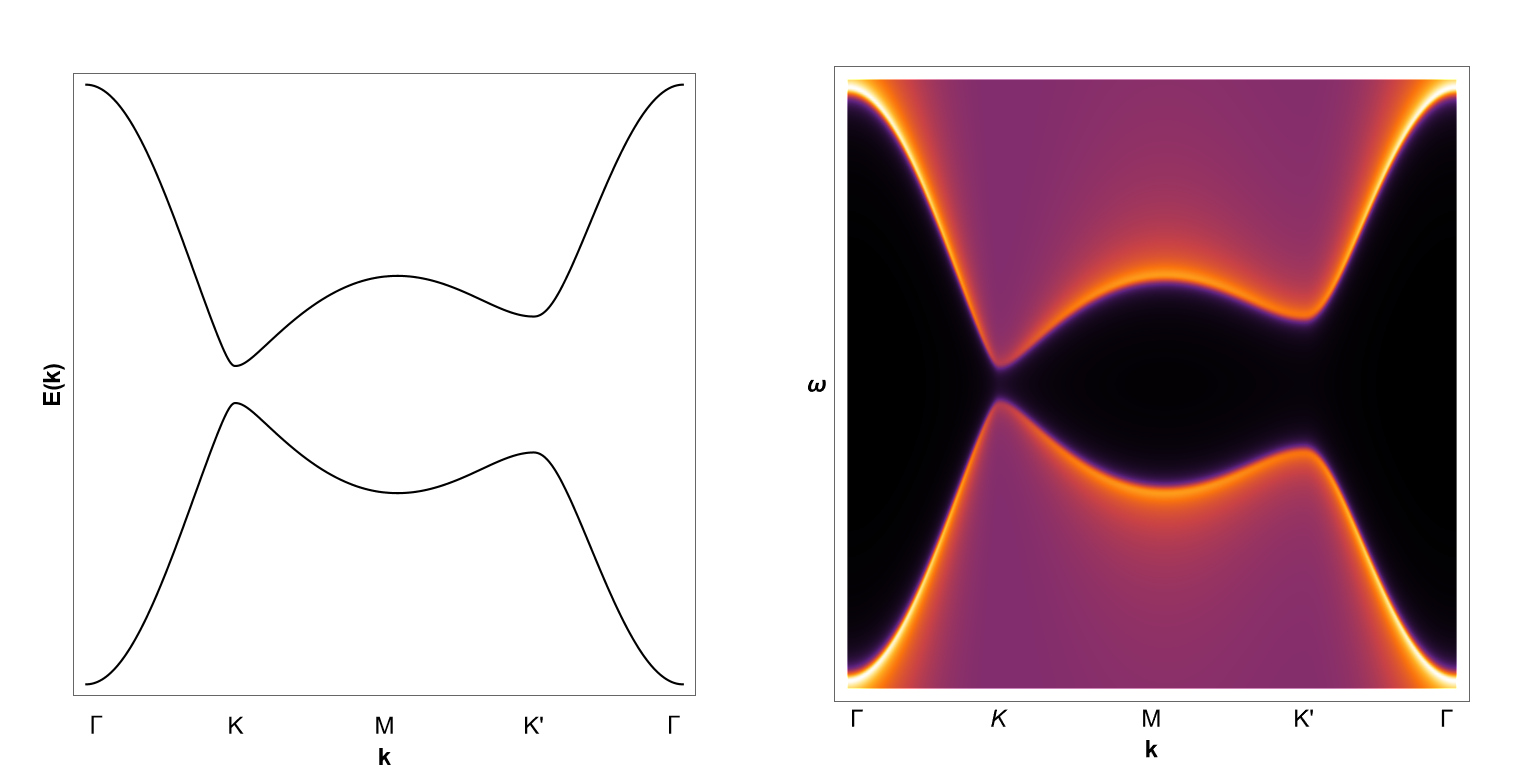}}
	\caption{Haldane model's dispersion relation(DR) and spectral density(SD). Similar to the graphene, DR is also well-described in SD.} \label{fig:Haldane}
\end{figure}

	\paragraph{Haldane model}
	Again, the Haldane model's tight-binding Hamiltonian is given by
\begin{align}
	h_{\text{TB}}(\boldsymbol{k})=&\,2t_2\cos(\phi)\left(2\cos(\tilde{k}_x)\cos(\tilde{k}_y)+\cos(2\tilde{k}_y)\right)I_2-t\left(1+2\cos(\tilde{k}_x)\cos(\tilde{k}_y)\right)\sigma_1 \nonumber \\
	&-2t\sin(\tilde{k}_x)\cos(\tilde{k}_y)\sigma_2+\left(\lambda_{v}+2t_2\sin(\phi)\left(2\cos(\tilde{k}_x)\sin(\tilde{k}_y)-\sin(2\tilde{k}_y)\right)\right)\sigma_3.
\end{align}
By substituting the above Hamiltonian into Green's function, we can depict the dispersion relation and spectral density in Figure \ref{fig:Haldane}.
Similar to the graphene case, in Figure \ref{fig:Haldane}, the spectral density edge location matches the Haldane model's tight binding result shown in the left figures. However, here also, we find that degrees of freedom(DoF) in SD spread out the whole region of $(k,\omega)$ within a time-like region, the same as the previous example.

	\section{Fermion bulk mass as a measure of Coupling strength}
\begin{figure}[H]
	\centering
	\captionsetup{justification=centering}
	\subfloat[$m\simeq0.5$]{\includegraphics[width=5.5cm]{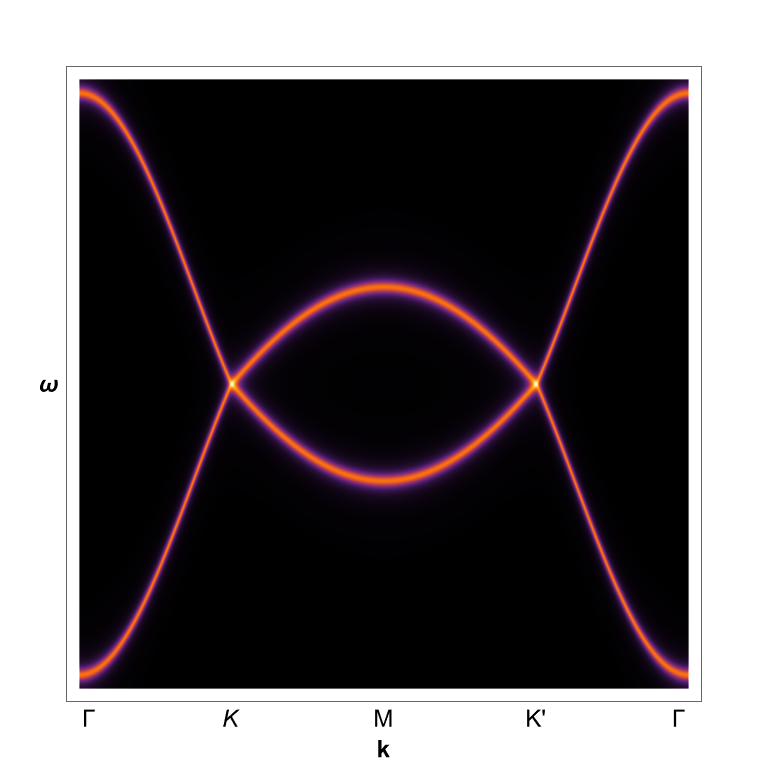}}
	\subfloat[$m=0.25$]{\includegraphics[width=5.5cm]{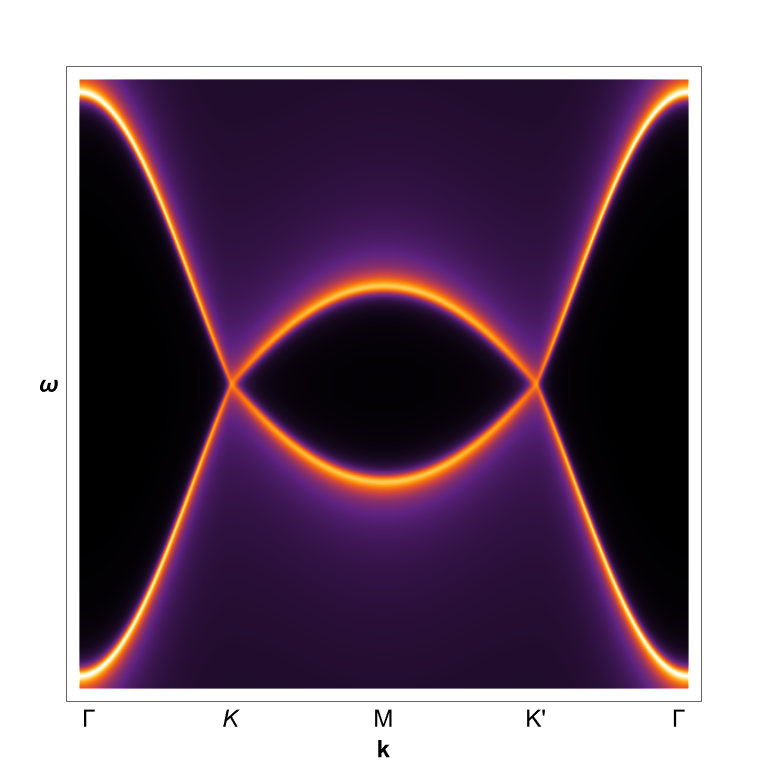}}
	\subfloat[$m=0$]{\includegraphics[width=5.5cm]{./fig/Graphene_m0}}
	\caption{Graphene's SD changed by $m$ with $t=1$, $t_2=0$ and $\lambda_{v}=0$. As we change the value of $m$ from $\frac{1}{2}$ to $0$, there is a transition from a simple pole to a branch-cut type pole.} \label{fig:Graphenemch}
\end{figure}

	This section considers the fermion mass, denoted as $m$, as an indirect control parameter influencing the ee interaction. This is based on considering an anomalous dimension $\gamma(m)$ as introduced in \cite{Cubrovic:2009ye}. Analyzing the result derived from Equation \eqref{eq:analyticGF}, we draw attention to the power of the denominator given by $\frac{1}{2}-m$. As $m$ decreases from $\frac{1}{2}$ to $0$, the SD undergoes a transition from a simple pole to a branch-cut singularity, as illustrated in Figure \ref{fig:Graphenemch}.
It is pertinent to note that in weakly interacting systems, a simple pole is typically observed.
As the interaction strength grows, the pole's residue diminishes, the band structure broadens, and eventually, the residue vanishes, entering the non-fermi liquid regime. One may ask whether the spectral function becomes completely fuzzy or it can maintain some singularity. Our calculation answers the question by confirming the second possibility.  
It shows that the strongly correlated systems can exhibit a branch-cut type singularity, although the system completely loses the quasiparticle's character. Next, we define the anomalous dimension as the difference between holographic scaling dimension $\Delta$ and free fermion's  $\Delta_0$ in $(2+1)$ dimension:
\begin{align}
	\gamma(m):=\Delta-\Delta_0=\left(\frac{3}{2}-m\right)-1=\frac{1}{2}-m.
\end{align}
Of particular interest are the cases when $m=0$ and $m=\frac{1}{2}$, yielding the following result:
\begin{align}
	\gamma(m)=
	\begin{cases}
		\frac{1}{2} \quad \text{for} \quad m=0 \\
		0 \quad \text{for} \quad m=\frac{1}{2}
	\end{cases}.
\end{align}
For $m=\frac{1}{2}$,  the anomalous dimension($\gamma$) is zero, suggesting the  Fermi liquid phase \cite{Cubrovic:2009ye}. Conversely, when $m=0$, $\gamma$ reaches its maximum value, implying a maximum effect of the strong correlation. Therefore, we use  $m$ as the control parameter for the ee interaction.
	
	\section{Conclusion}
	In this work, we introduced a lattice encoding method by embedding the tight-binding Hamiltonian into the AdS Dirac equation. The approach, grounded in the gauge gravity duality, facilitates constructing Green's function with the lattice effect encoded as well as the strong electron-electron interaction effects. The efficacy of our proposed embedding method is underscored through a comparison of dispersion relation (DR) and spectral density (SD). Furthermore, this analysis yields additional insights, such as intensity and width, not encompassed by band structure alone.
	
	By comparing the DR and SD depicted in Figures \ref{fig:Graphene} and \ref{fig:Haldane}, it is evident that the shape of the DR curve is respected in the SD. This observation substantiates the validity of our employed embedding method. Additionally, our approach yields other information, such as intensity and decay width. Notably, our results deviate from those typical of weakly interacting systems in that our Green's function shows branch cut singularity supporting the non-fermi liquid behavior of the electron fluid. 
	
	The fermion mass ($m$) interpretation is elucidated by introducing an anomalous scaling dimension. This conceptualizes $m$ as an indirect parameter influencing ee interactions, revealing a relationship between fermion mass and the system's phase.
	
	There are prospects for further research. Extending the methodology to 3- and 4-bands with spin-orbit coupling will enable the exploration of topological properties. Additionally, applying the technique to holographic p- and d-wave superconductors with anisotropic geometry helps to understand the strongly correlated system.
	
	\acknowledgments
	This work is supported by Mid-career Researcher Program through the National Research Foundation of Korea grant No. NRF-2021R1A2B5B02002603,  RS-2023-00218998 and NRF-2022H1D3A3A01077468. 
	We also thank the APCTP for the hospitality during the focus program, “Quantum Matter and Entanglement with Holography”, where part of this work was discussed.
	 
	 \appendix
\section{Analytic calculation of the flow equation}
This section shows the calculation of analytic Green's function via flow equation. The equation is given by
\begin{align}
	\partial_r\xi(r)+(\omega-h_0-\boldsymbol{h} \cdot \boldsymbol{\sigma})+\frac{2m}{r}\xi(r)+\xi(r)(\omega-h_0+\boldsymbol{h} \cdot \boldsymbol{\sigma})\xi(r)=0.
\end{align}
After, diagonalization of $\boldsymbol{h}\cdot\boldsymbol{\sigma}$, the above equation is written as
\begin{align}
	\partial_r\tilde{\xi}(r)+(\omega-h_0-|\boldsymbol{h}|\sigma_3)+\frac{2m}{r}\tilde{\xi}(r)+\tilde{\xi}(r)(\omega-h_0+|\boldsymbol{h}|\sigma_3)\tilde{\xi}(r)=0,
\end{align}
where
\begin{align}
	\tilde{\xi}(r)=U^{-1}\,\xi(r)\,U, \quad U=\frac{\sqrt{|\boldsymbol{h}|^2-h_3^2}}{\sqrt{2|\boldsymbol{h}|}}
	\begin{pmatrix}
		\frac{\sqrt{|\boldsymbol{h}|+h_3}}{h_1+ih_2} & -\frac{\sqrt{|\boldsymbol{h}|-h_3}}{h_1+ih_2} \\
		\frac{1}{\sqrt{|\boldsymbol{h}|-h_3}} & \frac{\sqrt{1}}{\sqrt{|\boldsymbol{h}|+h_3}}
	\end{pmatrix}.
\end{align}
Now, take the following ansatz:
\begin{align}
	\tilde{\xi}(r)=diag(\mathcal{F}_{+}(r),\mathcal{F}_{-}(r)).
\end{align}
Then, the equation can be represented as
\begin{align}
	\partial_r\mathcal{F}_{\pm}(r)+(\omega-h_0\mp|\boldsymbol{h}|)+\frac{2m}{r}\mathcal{F}_{\pm}(r)+(\omega-h_0{\pm}|\boldsymbol{h}|)\mathcal{F}_{\pm}(r)^2=0.
\end{align}
With the horizon value $$\tilde{\xi}(r_H)=iI_2,$$ the solution is given by
\begin{align}
	\mathcal{F}_{\pm}(r)=\frac{\omega-h_0\mp|\boldsymbol{h}|}{\sqrt{\boldsymbol{h}^2-(\omega-h_0)^2}}\frac{K_{m+\frac{1}{2}}(\sqrt{\boldsymbol{h}^2-(\omega-h_0)^2}\,r)}{K_{m-\frac{1}{2}}(\sqrt{\boldsymbol{h}^2-(\omega-h_0)^2}\,r)}.
\end{align}
Now, recast to $\xi(r)$ as follows:
\begin{align}
	\xi(r)=\frac{1}{\sqrt{\boldsymbol{h}^2-(\omega-h_0)^2}}\frac{K_{m+\frac{1}{2}}(\sqrt{\boldsymbol{h}^2-(\omega-h_0)^2}\,r)}{K_{m-\frac{1}{2}}(\sqrt{\boldsymbol{h}^2-(\omega-h_0)^2}\,r)}
	\begin{pmatrix}
		\omega-h_0-h_3 & -h_1+ih_2 \\
		-h_1-ih_2 & \omega-h_0+h_3
	\end{pmatrix}.
\end{align}

\section{Choice of $h_0$ position}
First, we rewrite the two kinds of embeddings as follows:
\begin{align}
	&\textbf{Standard Embedding}\text{:} \quad \left(\omega+q A_t-h_0\right)-\mathcal{G}\boldsymbol{h}\cdot\boldsymbol{\alpha}, \\
	&\textbf{Other possibility}\text{:} \quad \left(\omega+q A_t-\mathcal{G}h_0\right)-\mathcal{G}\boldsymbol{h}\cdot\boldsymbol{\alpha}.
\end{align}
The critical difference between the two embeddings is the presence of $\mathcal{G}$, which cannot be distinguished in the zero temperature limit ($\mathcal{G}(r)=1$). Especially the flow equation shows a difference more evidently:
\begin{align}
	\textbf{Standard}\text{:} \quad &\partial_r\xi(r)+(\frac{\omega+q A_t-h_0}{f}-\frac{\boldsymbol{h}\cdot \boldsymbol{\sigma}}{\sqrt{f}} ) \\
	&+\frac{2m}{r\sqrt{f}}\xi(r)+\xi(r)(\frac{\omega+q A_t-h_0}{f}+\frac{\boldsymbol{h} \cdot \boldsymbol{\sigma}}{\sqrt{f}})\xi(r)=0, \\
	\textbf{Other}\text{:} \quad &\partial_r\xi(r)+(\frac{\omega+q A_t-h_0\sqrt{f}}{f}-\frac{\boldsymbol{h}\cdot \boldsymbol{\sigma}}{\sqrt{f}} ) \\
	&+\frac{2m}{r\sqrt{f}}\xi(r)+\xi(r)(\frac{\omega+q A_t-h_0\sqrt{f}}{f}+\frac{\boldsymbol{h} \cdot \boldsymbol{\sigma}}{\sqrt{f}})\xi(r)=0.
\end{align}
As we change the   {standard embedding }   to another possibility, there is an extra deformation function in front of $h_0$ :  
$	h_0 \, \rightarrow \, h_0\sqrt{f(r)}$, adding a reason to choose the standard one. 

To see what happens and to kill another possibility, we compare the spectral density of both embeddings; we can elucidate the role of $\sqrt{f(r)}$.
\begin{figure}[H]
	\centering
	\captionsetup{justification=centering}
	\subfloat[$\textbf{Choice 1}$]{\includegraphics[width=5.5cm]{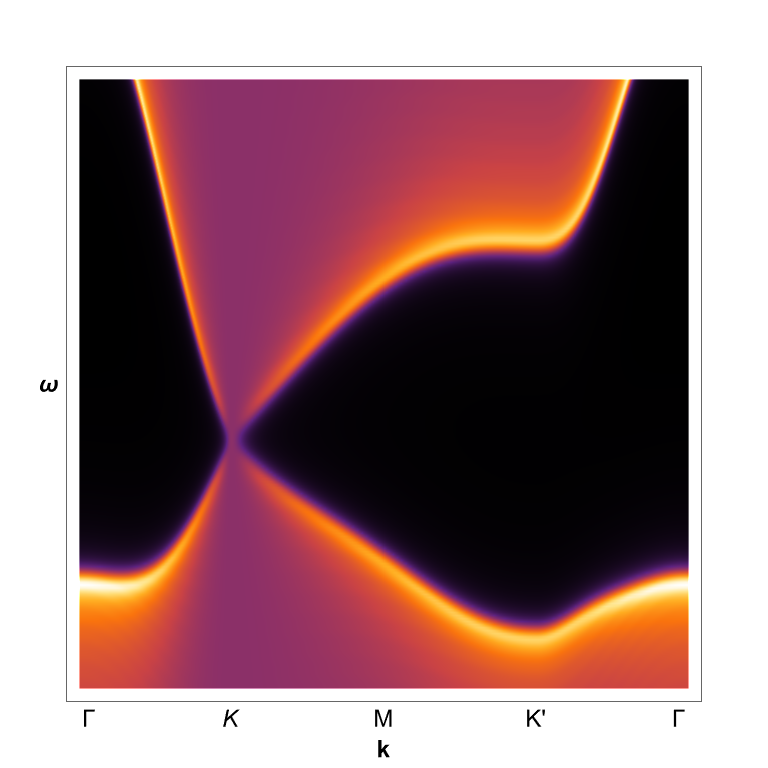}}
	\subfloat[$\textbf{Choice 2}$]{\includegraphics[width=5.5cm]{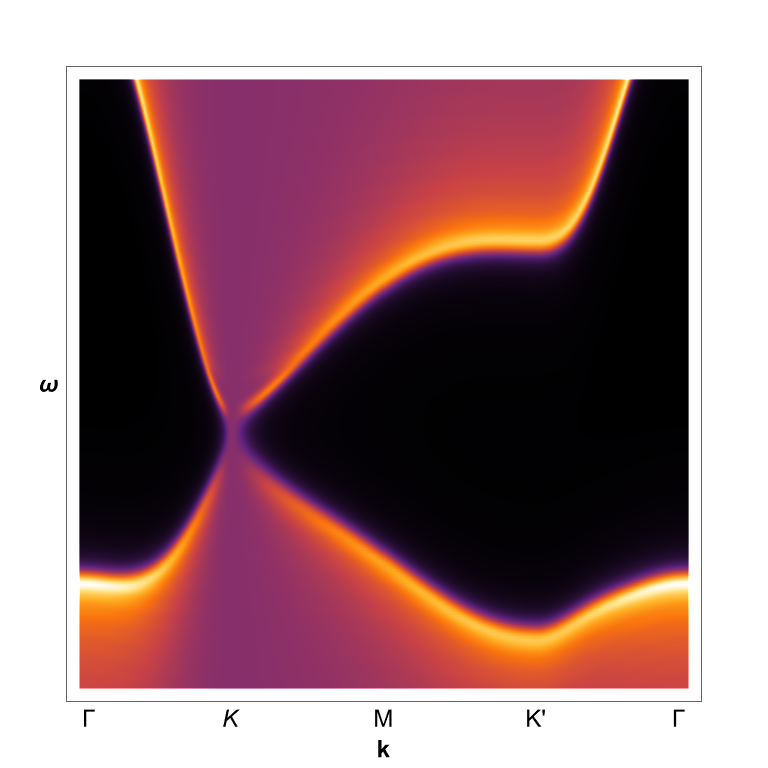}}
	\caption{Spectral density of Haldane model under the $(\phi,\frac{\lambda_{v}}{3\sqrt{3}t_2})=(\frac{\pi}{4},\frac{1}{\sqrt{2}})$ with two different choices. Near the Dirac cone, a bending effect occurs due to the power of $f(r)$.} \label{fig:choice}
\end{figure}
As we can see in Figure \ref{fig:choice}, there is no remarkable difference in the overall shape of spectral density. However, near the K-point, the standard embedding respects the Dirac cone. The difference comes from the different power of $f(r)$  in front of $h_0$. We expect that the significance of our choice will become evident when comparing our method with experimental data, such as Angel-resolved photoemission spectroscopy(ARPES).

\bibliographystyle{jhep}
\bibliography{TBH_ref.bib}

\end{document}